\documentstyle[epsfig,12pt]{article}
\def\beq{\begin{equation}}
\def\eeq{\end{equation}}
\def\bea{\begin{eqnarray}}
\def\eea{\end{eqnarray}}
\def\bq{\begin{quote}}
\def\eq{\end{quote}}

\def\NP{{\it Nucl. Phys.} }

\def\gappeq{\mathrel{\rlap {\raise.5ex\hbox{$>$}}
{\lower.5ex\hbox{$\sim$}}}}

\def\lappeq{\mathrel{\rlap{\raise.5ex\hbox{$<$}}
{\lower.5ex\hbox{$\sim$}}}}

\begin{document}
\pagestyle{empty}
\begin{flushright}
{CERN-TH/97-224}\\
{DEMO-HEP-97/06}\\
{PSI-PR-97-26}\\
\end{flushright}

\vspace*{5mm}
\begin{center}
{\bf ON THE DETERMINATION OF $M_W$ AND TGCs IN $W$-PAIR PRODUCTION}\\
{\bf USING THE BEST MEASURED KINEMATICAL VARIABLES}
\\
\vspace*{1cm} 
{\bf F.A. Berends} \\
\vspace{0.3cm}
Theoretical Physics Division, CERN \\
CH - 1211 Geneva 23 \\
and \\
Instituut Lorentz, University of Leiden, P.O. Box 9506,\\
2300 RA Leiden, The Netherlands \\
\vspace*{0.5cm} 
{\bf C.G.~Papadopoulos} \\
\vspace*{0.3cm}
Institute of Nuclear Physics, NCSR `Democritos',\\
15310 Athens, Greece \\
\vspace*{0.5cm} 
and \\
\vspace*{0.5cm} 
{\bf R. Pittau} \\
\vspace*{0.3cm}
Paul Scherrer Institute, CH - 5232 Villigen-PSI\\
\vspace*{1cm}  
{\bf ABSTRACT} \\ \end{center}
\vspace*{5mm}
\noindent
A study is made of the feasibility of maximum likelihood fits to determine
$M_W$ and triple gauge boson couplings using only those experimental
kinematical variables that are well measured. A computational tool to
calculate theoretical probabilities for those kinematical variables is
discussed and then applied to samples of unweighted events produced by an
event generator. Detailed results on the $M_W$ determination for
semileptonic final states in $W$-pair production show the feasibility of
the method. For TGCs one result is presented as an illustration.

\vspace*{1cm}

\begin{flushleft}
CERN-TH/97-224 \\
DEMO-HEP-97/06\\
PSI-PR-97-26\\
September 1997
\end{flushleft}

\vfill\eject

\setcounter{page}{1}
\pagestyle{plain}

At LEP 2, measurements of $W$-pair production and the subsequent decay of
the $W$-bosons are primarily used to determine the $W$-mass and the triple
gauge boson coupling (TGC). The theoretical description of the four
fermions produced in $W$-pair production leads to natural variables for
these measurements: invariant masses $M_1, M_2$ of the $W^\pm$ decay
products for the $M_W$ determination and a set of five angles for the TGC
determinations. The latter are the production angle $\Theta$ of $W^+$ in
the laboratory system and the decay solid angles $\Omega_1, \Omega_2$ of
an antifermion or fermion in the rest frames of a $W^+$ or $W^-$. On the
other hand, from the experimental point of view the well-measured
variables are energies and angles of charged leptons and angles of jets,
all in the laboratory system. The experimental resolution of these
variables is also well known. The set of these variables will henceforth be
denoted as $\{\phi\}$. To this set one might want to add the total 
hadronic energy as an additional variable.
Momenta of photons, mainly caused by initial-state
radiation (ISR), are also usually not accurately known experimentally. So
the reconstruction of the theoretically natural variables from the
well-measured experimental quantities poses a problem. One often has to
rely on some input from event generators for four-fermion production (for
a review on these generators, see~\cite{aaa}). Detailed accounts of a
direct reconstruction method for the $M_W$ determination~\cite{bb} and a
maximum likelihood method for the determination of TGCs~\cite{cc} have
been given in the final report of the Workshop on Physics at LEP 2.

Besides the complication of reconstructing natural variables, the maximum
likelihood method for TGCs poses another problem. The theoretical differential
cross-section with respect to the above-mentioned five angles is a natural 
quantity when the TGC effects are evaluated in a zero-width approximation
and without taking ISR into account.  When these approximations 
are not made,
the evaluation of this five-dimensional differential
cross-section, or in other words the probability $P(\Theta , \Omega_1,
\Omega_2, \{\alpha\})$ to find an event with $\Theta, \Omega_1, \Omega_2$
for a set $\{\alpha\}$ of anomalous TGCs, requires a fourfold
integration. In fact, one has to integrate over the invariant masses $M_1,
M_2$ and over the arguments $x_1, x_2$ of the $e^\pm$ structure
functions for the ISR (for the case of ISR in $W$-pair production, 
see~\cite{dd}).

This means that an efficient integration over these variables should be
done. Recently, the strategy of the TGC determination~\cite{ee} has been
discussed in a working group at the Oxford LEP 2 Workshop. 
The above-mentioned problems of reconstructing natural variables and of
required integrations led to a more general question: Can one successfully
determine anomalous couplings $\{\alpha\}$ from a maximum likelihood fit,
using for every event $i$, characterized by the experimentally
well-determined values $\{\phi_i\}$ of the kinematical variables
$\{\phi\}$, a theoretical probability $p(\{\phi_i\},\{\alpha\})$? In order
to answer this question, one first needs an efficient way to calculate
for an input $\{\alpha\}$ and $\{\phi_i\}$ the multidifferential
cross-section $d\sigma / d\phi$ and secondly one has to show that for
variables $\{\phi\}$ one can actually determine $\{\alpha\}$ by a  maximum
likelihood fit.

It is the purpose of this letter to study the feasibility of this
approach. Although strictly speaking only the TGC strategy was discussed
in~\cite{ee}, it is clear that the maximum likelihood method could also be
investigated for the $M_W$ determination. In fact, the methods to calculate\\ 
$p(\{\phi_i\},\{\alpha\})$ for the TGC set  $\{\alpha\}$ or 
$p(\{\phi_i\}, M_W)$ for a mass determination, are the same. The only
difference is that for the first case one needs the matrix element as a
function of 
$\{\alpha\}$, whereas in the second case only the Standard Model matrix
element is required. Thus the building of the analysing tool for the
extraction of $M_W$ or TGCs is in essence the same. Whether the tool
gives a good determination of $M_W$ or certain TGCs has to be established
by analysing data from event generators. In this letter results for the
$M_W$ determination will be presented, whereas the detailed TGC
results will be
deferred to a later paper~\cite{else}. The reasons to focus now on $M_W$ are
threefold. In the first place, the maximum likelihood method for $M_W$
offers an alternative to the reconstruction method. Secondly, it is the
simplest application of the developed computer program. Thirdly, the
variables $\{\phi\}$ and the integration algorithms are the same for the
$M_W$ and TGC determination.

In order to construct a tool for the evaluation of $p(\{\phi_i\}, M_W)$
one has to choose the set $\{\phi\}$ of accurately measured variables.
Although one can always consider more sets $\{\phi\}$,  the
following choices seem reasonable in practice~\cite{ee} for different 
four-fermion final states:

\begin{enumerate}
\item Semileptonic case: $q_1q_2\ell\nu$ \\
1a $\{\phi\} = (E_\ell, \Omega_\ell, \Omega_{q_1}, \Omega_{q_2})$ \\
1b $\{\phi\} = (E_\ell, \Omega_\ell, \Omega_{q_1}, \Omega_{q_2}, E_h)$, where
$E_h$ is the total energy of the jets.
\item Purely hadronic case: $q_1q_2q_3q_4$ \\
$\{\phi\} =\{\Omega_{q_1}\Omega_{q_2}\Omega_{q_3}\Omega_{q_4}\}$.
\item Purely leptonic case: $\ell_1 \nu_1 \ell_2 \nu_2$\\
$\{\phi\} = (E_{\ell_1}, \Omega_{\ell_1}, E_{\ell_2},\Omega_{\ell_2}\}$
\end{enumerate}

Since eight variables determine an event when no ISR is present, sets 1a
and 3 would require one or two integrations, cases 1b  and 2 none. 
Including ISR adds two integrations. When jets cannot be assigned to
specific quarks a folding over the various possibilities should be
included.

In order to make consistency checks, two programs have been developed to
calculate \\
$p(\{\phi_i\}, M_W)$, one program using parts of the event generator
ERATO~\cite{ff}, another similarly based on \mbox{EXCALIBUR}~\cite{ggg}. Details
will be published elsewhere~\cite{else}. For $\{\phi\}$ from set 1a, the program
calculates
\beq
p(\{\phi_i\}, M_W) = {1\over\sigma_{tot}}~~
{d\sigma\over dE_\ell d\Omega_\ell d\Omega_{q_1}d\Omega_{q_2}}~.
\label{one}
\eeq
When the experimental measurement range on $\{\phi\}$ is restricted,
$\sigma_{tot}$ should be calculated with the same restriction. For the
other sets 
$\{\phi\}$ the analogous multidifferential cross-sections are used and
calculated in the ``analyser" program.

Now we want to show that the maximum likelihood determination of $M_W$
works in the semileptonic case. To this end we generate 1600
unweighted events at a center-of-mass energy of 190 GeV,
with either ERATO or EXCALIBUR, for a known input value
\beq
M_W = 80.230 ~{\rm GeV}~.
\label{two}
\eeq
With the analyser we calculate
\beq
\log L = \sum_i \log p_i(M_W)~,
\label{three}
\eeq
where $i$ runs over the 1600 events and
\beq
p_i (M_W) = p(\{\phi_i\},M_W )~.
\label{four}
\eeq
For the $W$-mass parameter $M_W$ nine different values were chosen,
leading to nine data points for $\log L$, the logarithm of the likelihood
function. The integration needed to calculate $p_i(M_W)$ leads to a
small error on those data points. A parabola is fitted to these points
from which the reconstructed $W$-mass $M_R$ is obtained with an error
$\Delta M_R$.

An example of this procedure is shown in Fig. 1. The 1600 events were
generated in the whole phase space with a CC3 matrix element 
(containing only the three signal
diagrams~\cite{dd}) including the ISR. Also the analyser contained CC3
diagrams and ISR. The reconstructed mass is 80.238 $\pm$ 0.049. So 
within the errors (mainly determined by the number of experimental points,
i.e. 1600) the correct $W$-mass
 has been recovered. The variable set 1a has been used in this analysis.

In order to get more insight in the reliability of the maximum likelihood
procedure to find $M_R$ and to study some physics effects, three different
data sets of 1600 events were used, which were analysed in different ways.
We label the various data sets with $j = 1,2,3$. All three include ISR,
the first one is based on CC3, the second on CC10 and the last one on
CC20. So the pure signal case is generated, the muon
case with all diagrams and the electron case with all diagrams. 
The CC3 and CC10 events are produced in the whole phase space, while, for the
CC20 set, the outgoing $e^-$ has a $10^\circ $ cut from the beam.

The
various analysers are denoted by $i = 1,5$ and are respectively CC3, CC3 +
ISR, CC10 + ISR, CC20 + ISR, CC3 + ISR + folding. In the last case the
average cross-section is used with the two possible quark assignments. The
results for a number of $(i,j)$ combinations are given in Table 1 for
case 1a and in Table 2 for case 1b. The entries are the reconstructed
masses.

{From} Table 1, the following conclusions can be drawn. Comparing cases (1,1)
and (2,1) shows that neglect of ISR in the analysis causes a shift of 
$269 \pm 49$ MeV. Comparing (2,1) and (5,1) shows that folding gives 
within the errors a well-reconstructed $M_R$.

Comparing cases (2,2) and (3,2) tells us that the background diagrams, in the
case the final state lepton is a muon, have no significant effect. 
An analysis with CC3 alone is sufficient, within the errors.
The same happens for the case the final state lepton is an electron
(see cases (2,3) and (4,3)). However, the latter statement depends 
on the error originating from just using 1600 events and on the phase 
space cuts. Here a $10^\circ$ cut from the beam is applied on the
outgoing $e^-$. 
Cutting less may give a larger background and a significant shift.

Similar conclusions can be drawn when the set 1b is used. The errors on
the reconstructed $M_R$ are better, since one less integration is needed.
Also the shift in $M_R$ from the ISR is smaller, now being 
$142 \pm 33$ MeV.

Variable set 3 is not as much efficient for the mass reconstruction.
This is due to the fact that a large part of the kinematical
information is actually missing. We have found that the determination 
error in this set of variables is twice as much as in the case 1a for the same
number of generated unweighted events. The purely hadronic case can also be
used. Since the folding complicates the calculation of $p_i$, we defer the
discussion of this case to a future paper~\cite{else} .

The analyser gives the probability as a function of a limited set of
kinematical variables $\{\phi\}$. It integrates over some undetermined
variables and ISR. Whereas in event generators one wants to generate
events with all information provided, also preferably ISR photon momenta,
the analyser only has to integrate over the ISR. It is known that the
structure function method gives a reliable integration over all emitted
ISR photon momenta. Because of this ISR integration in the analyser
program, the details of $p_T$ distributions of ISR photons are expected
to matter here less than for 
refined event generation, where they matter for the kinematics
of each event. It should be noted that the unweighted events used for
the analysis of Tables 1 and 2 are generated with ISR along either
beam direction. In order to test the effect of transverse photon
momenta of the ISR, another event sample generated by KORALW~\cite{ko}
was analysed.
The inclusion of transverse photon momenta did not give results
different from the ones in the tables, at least within the errors
of the analysis.
{From} this point of view, the maximum likelihood method for
determining $M_W$ may have an advantage over the direct reconstruction
method.

In conclusion, the maximum likelihood method for determining $M_W$ from
well-measured experimental variables, with well-known resolutions, seems to
offer an alternative to the direct reconstruction method.

Besides the incorporation of the ISR, the method also easily takes
into account the full set  of diagrams for each channel. Beforehand one
can test whether a CC3 + ISR analysis is sufficient, or 
whether background diagrams should be included.

Although further results concerning the full
set of TGC parameters, as well as the consideration of all
possible production channels, will be presented elsewhere~\cite{else},
we give here one indication of its feasibility.
We have performed an analysis based on
the $\alpha_{W\phi}$ parameter~\cite{cc}. More precisely
the 1600 events generated with CC3+ISR have been analysed in the
case 1a, with the result: 
\[  
\alpha_{W\phi} = 0.002 \pm 0.026\;\;.
\]
We also generated 1600 events with CC3 but without ISR and analysed
them using the set of variables described in 1b. In that case, 
where no ISR is present, we expect to have a more or less ideal situation,
since the full kinematical information is available. The 
fitted value is now 
\[  
\alpha_{W\phi} = 0.006 \pm 0.024\;\;.
\]
Both numbers agree very well with the results presented 
in~\cite{cc}, where in the so-called `ideal' case
a determination error of 0.018 has been found by fitting
a sample of 2600 events. 

Summarizing, the feasibility of the method has been 
demonstrated from a theoretical point of view. Further
studies with detector simulation would be required to
see the results of the method in an experimental
environment.
\vspace*{0.5cm}\\

\noindent {\large \bf Acknowledgements:}
Discussions with the members of the Oxford TGC working group
are gratefully acknowledged. The KORALW sample of events was
kindly provided to us by \mbox{Z. Was.}

\clearpage

\begin{figure}[h]
\center
\vskip 0cm
\hskip 0cm
\epsfig{figure=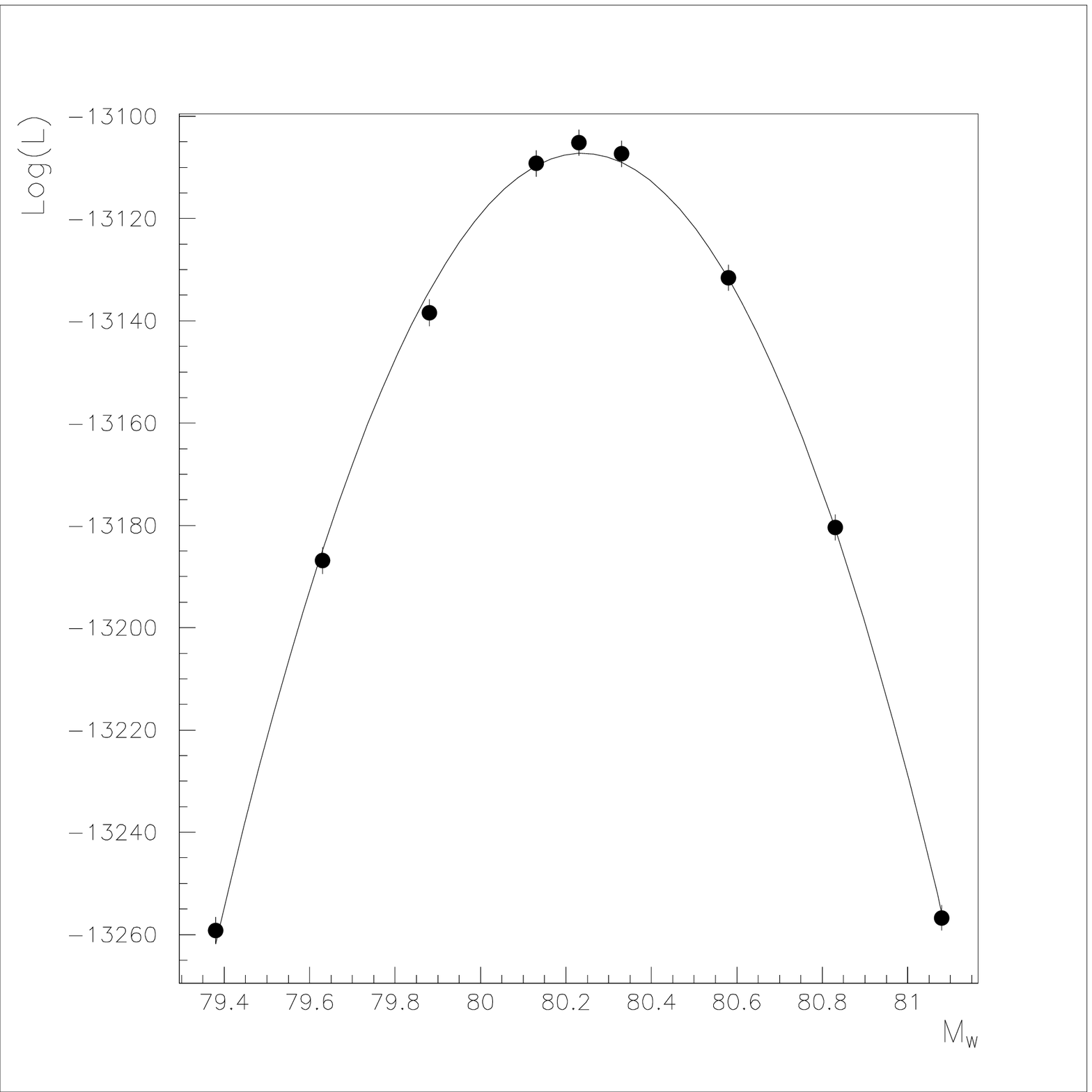,height= 6.in}
\caption{Example of likelihood curve for case $i= 2,~j= 1$
in Table 1.}
\end{figure}

\clearpage

\begin{table}
\begin{center}
\begin{tabular}{||l | c | c | c  ||}  \hline
$j = ~~~~\rightarrow$ & 1 & 2 & 3 \\
$i = ~~\downarrow$ & CC3 + ISR & CC10 + ISR & CC20 + ISR \\ \hline
1~~CC3 & 80.507 $\pm$0.045 & & \\
&  & & \\ \hline
2~~CC3 + ISR & 80.238 $\pm$0.049 & 80.224 $\pm$ 0.048 & 80.280 $\pm$ 0.047\\
&  &  & \\ \hline
3~~CC10 + ISR & & 80.277 $\pm$ 0.048 & \\
&&& \\ \hline
4~~CC20 + ISR & & &80.231 $\pm$ 0.049\\
&&&\\ \hline
5~~ CC3 + ISR & 80.249 $\pm$0.048 && \\
+ folding &  && \\ \hline
\end{tabular}
\end{center}
\caption{The reconstructed $M_R \pm \Delta M_R$ from three different
data samples of 1600 unweighted events $(j = 1,2,3)$. The analyser 
has been used for five different treatments of the cross-section $(i = 1,5)$. 
Sets $j= 1$ and $j= 2$ have no cuts, while, for set $j= 3$, the
outgoing $e^-$ has a $10^\circ$ cut from the beam.
The experimental variables are those of set 1a.}

\end{table}

\begin{table}
\begin{center}
\begin{tabular}{||l | c | c | c  ||}  \hline
$j = ~~~~\rightarrow$ & 1 & 2 & 3 \\
$i = ~~\downarrow$ & CC3 + ISR & CC10 + ISR & CC20 + ISR \\ \hline
1~~CC3 & 80.380 $\pm$0.032 & & \\
&  & & \\ \hline
2~~CC3 + ISR & 80.238 $\pm$0.033 & 80.238 $\pm$ 0.032 & 80.269 $\pm$ .032 \\
&  &  & \\ \hline
3~~CC10 + ISR & &80.240 $\pm$ 0.032 & \\
&&& \\ \hline
4~~CC20 + ISR & & & 80.240 $\pm$ 0.033 \\
&&&\\ \hline
5~~ CC3 + ISR & 80.239 $\pm$0.033 && \\
+ folding &  && \\ \hline
\end{tabular}
\end{center}
\caption{The reconstructed $M_R \pm \Delta M_R$ from three different 
data samples of
1600 unweighted events $(j = 1,2,3)$. The analyser has been used for
five different treatments of the cross-section $(i = 1,5)$. 
Sets $j= 1$ and $j= 2$ have no cuts, while, for set $j= 3$, the
outgoing $e^-$ has a $10^\circ$ cut from the beam.
The experimental variables are those of set 1b.}
\end{table}

\end{document}